\documentclass[prl,showpacs,twocolumn]{revtex4-1}
\usepackage{amssymb,amsmath}
\usepackage{natbib}
\usepackage{graphicx}

\defcitealias{Levine_etal98}{LRFB}

  \def \pe {\text{Pe}}
	 \renewcommand{\vec}[1]{\mathbf{#1}}

\begin{document}

\title{Screening, Hyperuniformity, and Instability in the Sedimentation of Irregular Objects}

\author{Tomer Goldfriend}
\email{goldfriend@tau.ac.il}
\affiliation{Raymond \& Beverly Sackler School of Physics and Astronomy, Tel Aviv
University, Tel Aviv 69978, Israel}

\author{Haim Diamant} 
\email{hdiamant@tau.ac.il} 
\affiliation{Raymond \& Beverly Sackler School of Chemistry, Tel Aviv
University, Tel Aviv 69978, Israel}

\author{Thomas A.\ Witten} 
\email{t-witten@uchicago.edu}
\affiliation{Department of Physics and James Franck Institute,
University of Chicago, Chicago, Illinois 60637, USA}

\date{\today}

\begin{abstract}
We study the overdamped sedimentation of non-Brownian objects of irregular shape using fluctuating hydrodynamics. The anisotropic response of the objects to flow, caused by their tendency to align with gravity, directly suppresses concentration and velocity fluctuations. This allows the suspension to avoid the anomalous fluctuations predicted for suspensions of symmetric spheroids. The suppression of concentration fluctuations leads to a correlated, hyperuniform structure. For certain object shapes, the anisotropic response may act in the opposite direction, destabilizing uniform sedimentation.  
\end{abstract}


\maketitle

Sedimentation, the settling of colloidal objects under gravity, is a fundamental and ubiquitous physical process whose details are still under debate (see reviews in Refs.~\cite{Ramaswamy2001,Gauzzelli&Hinch2011}). The related process of bed fluidization is widely used in reactors, filtration, and  water treatment~\cite{JohnsonHandBook}. Long-range hydrodynamic correlations among settling objects lead to complex many-body dynamics, exhibiting strong fluctuations and large-scale dynamic structures even for athermal (non-Brownian) objects with negligible inertia~\cite{Ham&Homsy88,Nicolai&Gauzzelli95,Segre_etal97,Tong&Ackerson98}. One of the key issues is the extent of velocity fluctuations of the sedimenting objects about their mean settling velocity. A famous prediction by Caflisch and Luke~\cite{Caflisch&Luke85} stated that the magnitude of velocity fluctuations of individual objects should diverge with system size. Over the years there has been evidence from theory and simulations both in favor of~\cite{Ladd96,Ladd97,Koch1994,Cunha_etal2002,Bergougnoux_etal2003,Mucha_etal2004} and against~\cite{Koch&Shaqfeh91,Levine_etal98,Ramaswamy2001,Brenner99} this prediction. Experimentally, the indefinite growth of velocity fluctuations with system size has not been observed~\cite{Nicolai&Gauzzelli95,Segre_etal97}. 

To resolve the Caflisch-Luke paradox, several screening mechanisms have been suggested: a characteristic screening length emerging from correlations between concentration fluctuations (the structure factor of the suspension)~\cite{Koch&Shaqfeh91}, e.g., as a result of stratification~\cite{Luke2000,Mucha_etal2004}; inertial effects~\cite{Brenner99}; side-wall effects~\cite{Brenner99}; and noise-induced concentration fluctuations~\cite{Levine_etal98,Ramaswamy2001}.

Earlier theories have considered symmetric objects, mostly
spheres. Spheroids~\cite{Koch&Shaqfeh89,Dahlkild2011}, rod-like
objects~\cite{Kumar&Ramarao91,Turney_etal95,Herzhaft&Guazzelli99,Metzger_etal2007,Saintillan_etal2005,Butler&Shaqfeh2002,Gustavsson&Tornberg2009},
and permeable spheres~\cite{Cichocki_etal2011}, were studied as
well. In various scenarios, including applications involving fluidized
beds, the suspensions contain objects of asymmetric shapes. In the
present work we address the sedimentation of a large class of
irregular objects which are self-aligning~\cite{Krapf_etal2009}. Under
gravity, in addition to settling, such an individual object aligns an
eigendirection with the driving force. This should be distinguished
from symmetric objects like rods, which align with flow
lines~\cite{Koch&Shaqfeh89,Herzhaft&Guazzelli99} rather than with
an external force of fixed direction. In general these objects are chiral
and thus also rotate about the force direction in a preferred sense of
rotation. Both the eigendirection and angular velocity are determined
by the object's geometry and mass
distribution~\cite{Krapf_etal2009}. The hydrodynamic pair-interactions
between self-aligning objects have been studied in
Refs.~\cite{Goldfriend_etal2015I,Goldfriend_etal2015II}. Unlike
spheres, the objects respond anisotropically to nonuniform flow. This
fact, as shown below, suppresses fluctuations for arbitrarily weak
inhomogeneity (unlike the case of spheres studied in
Ref.~\cite{Levine_etal98}).

We begin with a qualitative description of the effects studied here. Consider a suspension of objects sedimenting in a viscous fluid of viscosity $\eta$ under force $F$ in the $-z$ direction. The mean concentration is $c_0$.
Let us imagine a sinusoidal variation $c(x)$ about $c_0$, of wavelength $\lambda$, in the transverse $x$ direction, creating vertical slabs of heavier and lighter weights.
This creates a velocity variation, $U(x)$. To find the amplitude of this variation we balance the change in gravitational force with the change in viscous drag (per unit area of the slab), $c \lambda F\sim \eta U/\lambda$, resulting in $ U \sim c \lambda^2 F/\eta$. This indefinite increase of $U$ with $\lambda$ is a manifestation of the Caflisch-Luke problem. The relative velocity of the slabs creates a vorticity $\omega$ of order $ U / \lambda \sim  c \lambda F /\eta$. For spheres, this vorticity merely rotates the objects. Self-aligning objects, by contrast, are tilted away from their aligned state. Their misalignment, proportional to $\omega$, makes them glide in the $x$ direction with velocity $U_{\perp} \sim \gamma  a c \lambda F/\eta$, where $a$ is the size of the object and $\gamma$ a proportionality coefficient. The time derivative of concentration, arising from the gradient of flux, reads $\dot{c} \sim -c_0  U_{\perp} /\lambda = -(\gamma a c_0 F/\eta) c$. Now, if the coefficient $\gamma$ is positive, the response suppresses the inhomogeneity, whereas if it is negative the inhomogeneity is enhanced. This is a mechanism of either screening or instability. In addition, the independence of the last relation on $\lambda$ implies (for $\gamma>0$) a non-diffusive fast relaxation over large length scales. As shown below, this leads to a hyperuniform dynamic structure. By equating the diffusive and non-diffusive relaxation rates of a slab, $D\lambda^{-2}=\gamma a c_0 F/\eta$, where $D$ is the hydrodynamic diffusion coefficient, we find a typical wavelength above which hyperuniformity sets in, $\xi=[\gamma a c_0 F/(\eta D)]^{-1/2}$. Note that the mechanism just described does not work for concentration variations in the $z$ direction~\footnote{The situation for symmetric, non-aligning objects, such as spheroids is more complicated since the unperturbed object does not have a well-defined orientation with the force~\cite{Koch&Shaqfeh89}.}.       

To study these effects in more detail we use the framework of fluctuating hydrodynamics. Similar continuum approaches were used for spheres by Levine et al. (referred to hereafter as LRFB)~\cite{Levine_etal98}, and by Mucha et al.~\cite{Mucha_etal2004}. We consider an athermal inertia-less suspension. The system depends on the following parameters: the gravitational force on a single object, $F$; solvent viscosity $\eta$; characteristic size of the objects $a$; and mean concentration of objects $c_0$. In addition, a self-aligning object has an alignability parameter $\alpha$, giving the slowest relaxation rate of a mis-aligned orientation toward alignment, $\tau^{-1}_{\rm align}=\alpha F/(\eta a^2)$~\cite{Krapf_etal2009}. This parameter is derivable from the object's shape and mass distribution alone.

The stochastic response of the suspension is characterized by a phenomenological diffusion coefficient $D$, measurable in experiments~\cite{Ham&Homsy88}, and fluctuating object fluxes $\vec{f}(\vec{r},t)$, treated as a Gaussian white noise with variance $\langle f_i({\bf r},t) f_j ({\bf r}',t')\rangle = 2c_0 N \delta_{ij}\delta({\bf r}-{\bf r}') \delta(t-t')$~\cite{Ramaswamy2001,Levine_etal98}. The parameters $D$ and $N$, which originate in the complex many-body interactions excited by the force $F$ at each object, are in general anisotropic~\cite{Levine_etal98,Ham&Homsy88,Nicolai_etal95}. Yet, unlike LRFB model, the effects discussed below do not depend crucially on this anisotropy; we therefore neglect it for the sake of simplicity. 
In addition, $D$ may depend on the volume fraction due to shear-induced diffusion~\cite{Rusconi&Stone2008}. As this dependence was experimentally found to be weak~\cite{Ham&Homsy88,Nicolai_etal95}, we neglect it as well.

We employ the following three additional assumptions: (1) the suspension is dilute, having volume fraction $\varphi \ll 1$, such that direct interactions between the objects are negligible, and the hydrodynamic interaction is well described by its two leading multipoles.  (2) The suspension is non-Brownian, i.e., the thermal P\'eclet number $Fa/(k_B T) \gg 1$, where $k_B T$ is the thermal energy. However, there is no restriction on the {\it sedimentation} P\'eclet number, defined as $\pe \equiv F/(\eta D)$. (3) We assume strong alignability, i.e., that the rate of alignment $\tau^{-1}_{\rm align}$ is much faster than the interaction-induced vorticity, $\omega \sim 1/(\eta l^2)$, where $l\sim a\varphi^{-1/3}$ is the typical distance between objects. The resulting criterion, $\alpha \gg \varphi^{2/3}$, improves with dilution.

The advection-diffusion equation for the fluctuations of object concentration about $c_0$, $c(\vec{r},t)$, reads 
\begin{equation}
\partial_t c + \nabla\cdot((c+c_0){\bf U}) 
= D\nabla^2 c +\nabla\cdot{\bf f},
\label{eq:AdvDiff}
\end{equation}
where ${\bf U}$ is the objects' velocity fluctuation field about the mean settling velocity. The velocity fluctuation of the fluid surrounding the objects, $\vec{v}(\vec{r},t)$,  is described by an incompressible, overdamped Stokes flow, with force monopoles originating from concentration fluctuations~\footnote{Although a dipolar term of order $a$ should be included in Eq.~\eqref{eq:v} to have a consistent expansion to first order in $a$, this divergenceless term has no effect on the results.},
\begin{multline}
v_i({\bf r},t)=\int d^3r' G_{ij} ({\bf r}-{\bf r}') c({\bf r}',t) F_j({\bf r}') +O(a)
=\\
-F \int d^3r' G_{iz} ({\bf r}-{\bf r}') c({\bf r}',t)  +O(a).
\label{eq:v}
\end{multline}
Here $G_{ij}({\bf r})=(8\pi \eta r)^{-1} (\delta_{ij}+r_i r_j/r^2)$ is the Green's function of Stokes flow (the Oseen tensor)~\cite{Pozrikidis}. 

A point-like object ($a\rightarrow 0$) is merely advected by the flow, i.e., $\vec{U}=\vec{v}$. However, for nonzero $a$ the two velocities do not coincide. To leading order in $a$ they are bound to satisfy a relation of the form,
\begin{equation}
U_i=v_i +a \Phi_{ikj} \partial_j v_k + O(a^2).
\label{eq:Uv}
\end{equation}
The constant tensor $\vec{\Phi}$ depends on the objects' orientations and shapes, and is assumed to be independent of $c$~\footnote{The mean-field assumption of constant $\vec{\Phi}$ is valid provided that the motion of the isolated self-aligning object under external force and flow gradient can be characterized by axisymmetric hydrodynamic tensors~\cite{SM}.}. The difference between $\vec{U}$ and $\vec{v}$, and the fact that the effective response $\vec{\Phi}$ is anisotropic, lead to a new advective term in Eq.~\eqref{eq:AdvDiff}, which corresponds to an object flux with non-zero divergence, $\partial_i U_i = a \Phi_{ikj} \partial_i\partial_j v_k +O(a^2) \neq 0$. The second term in Eq.~\eqref{eq:Uv} is at the core of the present theory; the existence of asymmetry in $\Phi_{ikj}$, demanded phenomenologically for self-aligning objects, entails the effects described below (For spheres the second term in Eq.~\eqref{eq:Uv} vanishes, and the higher-order terms are divergence-less). The anisotropic response has two contributions: one from a direct translational response to shear flow, and the other due to the object's gliding response mentioned above~\cite{SM}.

We proceed by substituting Eqs.~\eqref{eq:v} and \eqref{eq:Uv} into Eq.~\eqref{eq:AdvDiff} and Fourier-transforming the resulting equation ($(\vec{r},t)\to(\vec{q},\omega)$).
This leads to
\begin{widetext}
\begin{multline}
 -i\omega \tilde{c}( \vec{q},\omega) +\frac{c_0 aF}{\eta}
\left( \gamma\frac{q^2_{\perp}}{q^2} + \bar{\gamma}\frac{q^4_{\perp}}{q^4} \right) \tilde{c}( \vec{q},\omega)+iF\int q_i\tilde{G}_{i3}(\vec{q}')\tilde{c}(\vec{q}-\vec{q}',\vec{\omega-\omega'})\tilde{c}(\vec{q}',\omega')d^3q' d\omega' \\
=-D q^2\tilde{c}( \vec{q},\omega)   -i\vec{q}\cdot\tilde{\vec{f}}( \vec{q},\omega),
\label{eq:FTAdvDiffN}
\end{multline}
\end{widetext}
where we used the fact that $q'_i\tilde{G}_{ij}(\vec{q}')=0$. We denote by $\perp$ the horizontal components $(x,y)$ of a vector. The coefficients $\gamma$ and $\bar{\gamma}$ are effective response parameters resulting from the response tensor $\vec{\Phi}$ (specifically, $\gamma=\Phi_{zzz}-\Phi_{z\perp\perp}-\Phi_{\perp\perp z}$, $\bar{\gamma}=\Phi_{\perp\perp z}+\Phi_{\perp z\perp}+\Phi_{z\perp\perp}-\Phi_{zzz}$). The second term in Eq.~\eqref{eq:FTAdvDiffN} corresponds to linear screening, which is nonzero for any wavevector $\vec{q}\nparallel \hat{\vec{z}}$. This term makes a simple perturbation theory in small concentration fluctuations valid, allowing us to neglect the third, nonlinear term that underlies the LRFB model. In addition, to facilitate the analysis, we omit the term proportional to $\bar{\gamma}$,  which does not affect the following calculations. We thus have
\begin{equation}
 \left(-i\omega + D q^2+ \gamma\frac{c_0 aF}{\eta}
 \frac{q^2_{\perp}}{q^2} \right) \tilde{c}( \vec{q},\omega)
=-i\vec{q}\cdot\tilde{\vec{f}}( \vec{q},\omega).
\label{eq:FTAdvDiff}
\end{equation}
By equating the diffusive and screening terms in Eq.~\eqref{eq:FTAdvDiff}, we obtain the characteristic length that we qualitatively inferred above,
\begin{equation}
\xi = \left( \frac{\gamma c_0  a F }{\eta D} \right)^{-1/2}
= a \gamma^{-1/2}\varphi^{-1/2} \pe^{-1/2}.
\label{eq:Xi}
\end{equation}

\begin{figure*}
\centerline{\resizebox{0.48\textwidth}{!}{\includegraphics{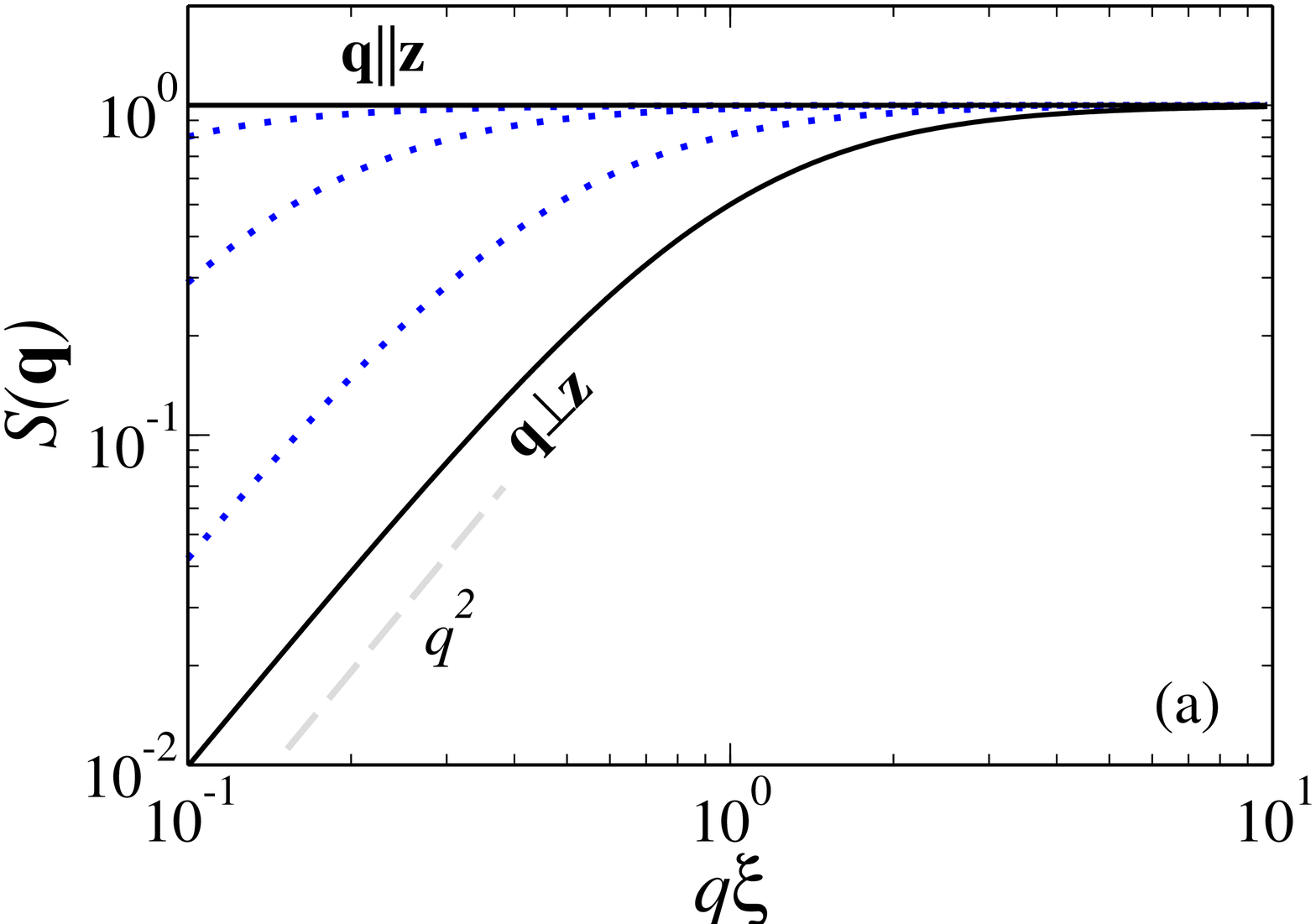}}
\hspace{0.5cm}
\resizebox{0.48\textwidth}{!}{
\includegraphics{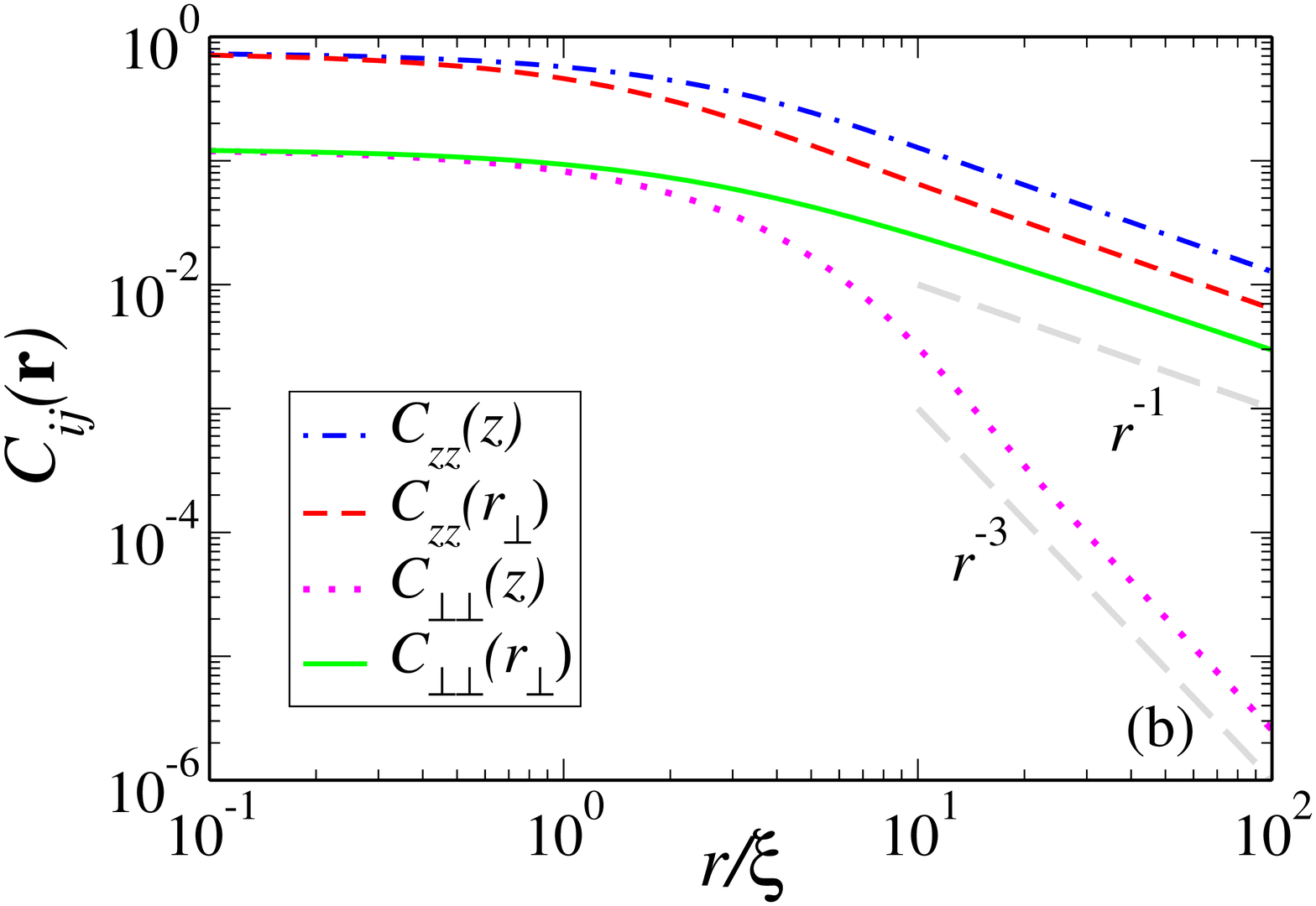}}}
\caption[]{(a) Static structure factor of the suspension (Eq.~\eqref{eq:Sq} with $N=D$). The blue dotted curves, which correspond to different $\vec{q}\nparallel \hat{\vec{z}}$, decay to zero at small $q$ as $\sim q^2$. 
(b) Normalized two-point velocity correlations, together with their asymptotic behavior (Eqs.~\eqref{eq:Cpp} and~\eqref{eq:Czz}).}  
\label{fig:fig1}
\end{figure*}

{\it Results}. We now summarize the main results, which are readily obtained from Eqs. \eqref{eq:v}--\eqref{eq:FTAdvDiff}. We begin with the expressions for the concentration and velocity correlation functions at steady state ($\omega\rightarrow 0$):
\begin{equation}
S(\vec{q})=\langle \tilde{c}(\vec{q},0)\tilde{c}(-\vec{q},0)\rangle =
\frac{N}{D}\frac{q^2}{q^2+\xi^{-2}(q_{\perp}/q)^2},
\label{eq:Sq}
\end{equation}
\begin{equation}
\langle \tilde{U}_i(\vec{q},0)\tilde{U}_j(-\vec{q},0)\rangle = \frac{N F^2}{D} 
\frac{\tilde{G}_{iz}(\vec{q}) \tilde{G}_{jz}(-\vec{q}) q^2}
{q^2+\xi^{-2}(q_{\perp}/q)^2},
\label{eq:UUq}
\end{equation}
where $S(\vec{q})$ is the static structure factor of the suspension. Figure~\ref{fig:fig1}(a) shows $S(\vec{q})$ along different directions of $\vec{q}$. The structure factor decays to zero at small $q$, as $q^2$, in all directions except $\hat{\vec{z}}$, where it is a constant at small $q$. Next, the velocity point-correlation functions are obtained by inverting back to real space and taking the limit $r\to 0$, 
\begin{multline}
\langle U^2_z(0)\rangle = 6 \langle U^2_{\perp}(0)\rangle =
\frac{3}{64}\varphi\frac{N}{D}\frac{\xi}{a}\left(\frac{F}{\eta a}\right)^2
\\
=\frac{3}{64} \frac{N}{D} \left(\frac{D}{a}\right)^2 \gamma^{-1/2} \varphi^{1/2} \pe^{3/2}.
\label{eq:sigmaU}
\end{multline}
Finally, we give the asymptotic expressions at large distances ($r\gg\xi$) for the two-point correlations in real space.
Finally, we give the asymptotic expressions at large distances ($r\gg\xi$) for the two-point correlations in real space.
For the concentration correlations we get
\begin{equation}
\frac{\langle c(0)c(r\hat{\vec{z}}) \rangle }{c_0 N/D \xi^3 }=
\frac{12}{\pi}\frac{\xi^5}{r^5}, \quad
\frac{\langle c(0)c(r\hat{\vec{r}}_{\perp}) \rangle}{c_0 N/D \xi^3 } =
\frac{\Gamma^2(5/4)}{2 \sqrt{2} \pi^2 }\frac{\xi^{5/2}}{r^{5/2}},
\label{eq:cc}
\end{equation}
where $\Gamma$ is the Gamma function. The weaker decay $\sim r^{-5/2}$ applies strictly within the $(x,y)$ plane.
For the velocity correlations we get
\begin{equation}
C_{\perp\perp}(r\hat{\vec{z}})=\frac{8\xi^3}{\pi r^3},
\qquad
C_{\perp\perp}(r\hat{\vec{r}}_{\perp})=\frac{\xi}{\pi r},
\label{eq:Cpp}
\end{equation}
\begin{equation}
C_{zz}(r\hat{\vec{z}})=\frac{4\xi}{\pi r},
\qquad
C_{zz}(r\hat{\vec{r}}_{\perp})= \frac{2\xi}{\pi r}, 
\label{eq:Czz}
\end{equation}
where $C_{ij}(\vec{r})\equiv\langle U_i(0)U_j(\vec{r})\rangle/\langle U^2(0)\rangle$.  Despite the emergence of the characteristic length $\xi$, the concentration and velocity correlations remain long-ranged, decaying algebraically with distance. In Fig.~\ref{fig:fig1}(b) we present the spatial correlations at steady state along with their asymptotic power laws.

\begin{figure*}
\centerline{\resizebox{0.4\textwidth}{!}{\includegraphics{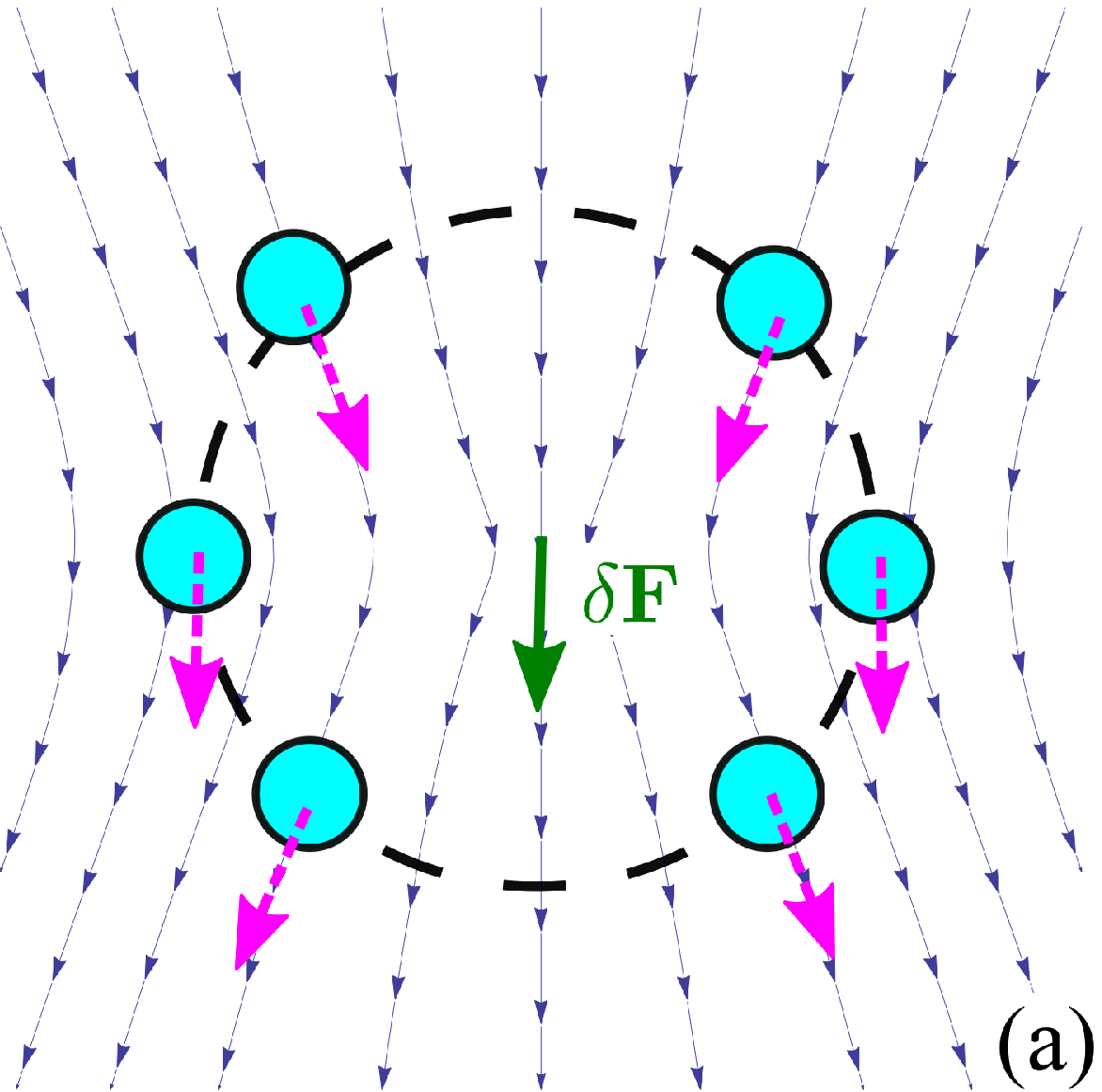}}
\hspace{0.5cm}
\resizebox{0.4\textwidth}{!}{
\includegraphics{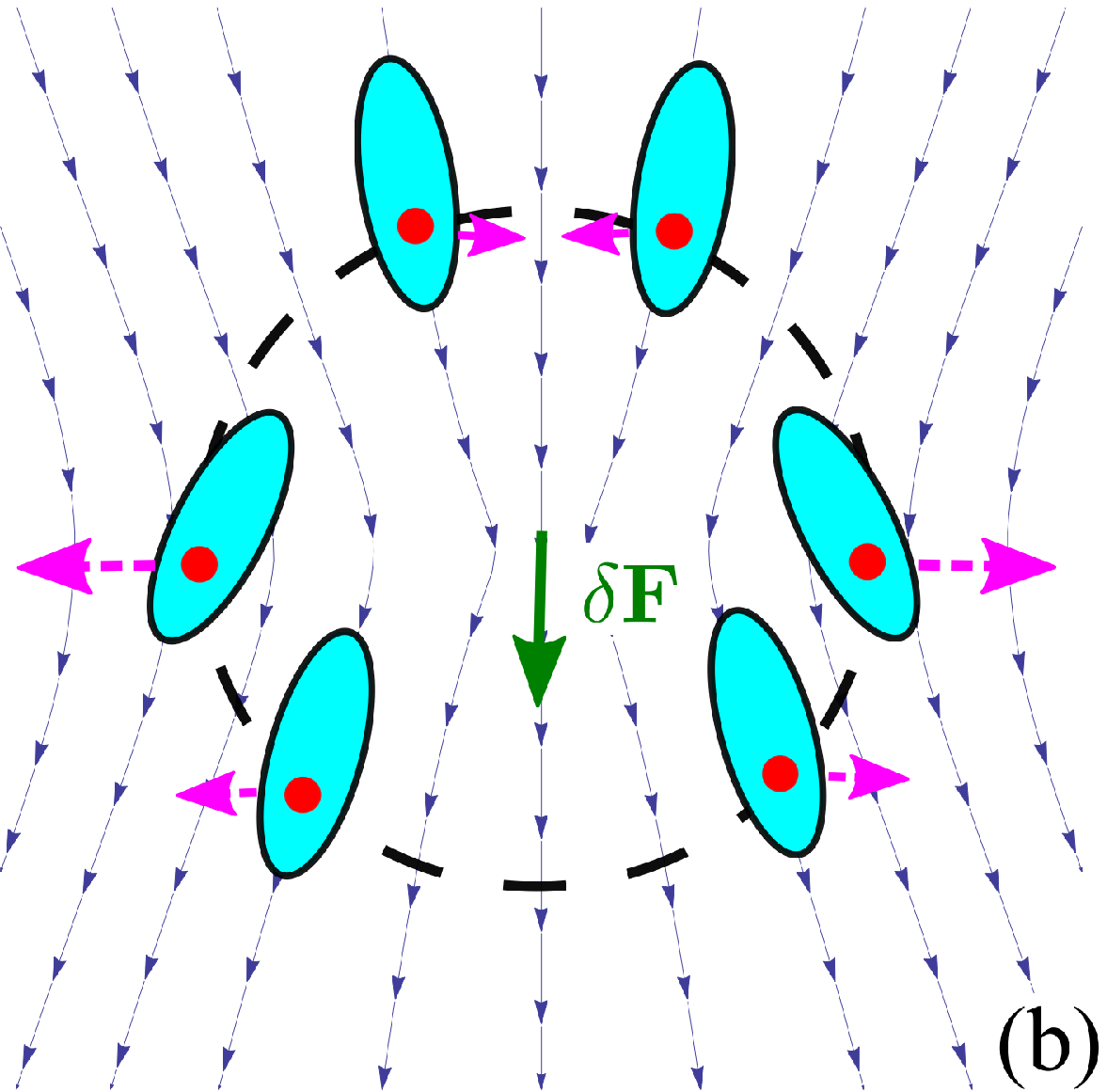}}}
\caption[]{Illustration of the mechanism regulating concentration fluctuations. A concentration fluctuation makes a force fluctuation $\delta \vec{F}$ (green solid arrow), which in turn creates a flow fluctuation (flow lines). 
(a) Spherical objects respond isotropically to this flow, with velocities (magenta dashed arrows) along the direction of the flow lines; thus, there is no net flux of objects into the small volume element around the fluctuation.
(b) Self-aligning spheroids have an anisotropic response, leading in this example to a total outflux of objects. The resulting flow of objects has a nonzero divergence, which reduces the concentration fluctuations.}  
\label{fig:fig2}
\end{figure*}

{\it Discussion.} Let us now discuss the consequences of these results. The velocity auto-correlation of an object is given, up to corrections of $O(a/\xi)$, by the point-correlation of Eq.~\eqref{eq:sigmaU}. From this expression we immediately see how the finite $\xi$ regularizes the velocity auto-correlations, thus removing the Caflisch-Luke problem~\cite{Caflisch&Luke85} for the irregular objects considered here. Indeed, in the limit $\gamma\rightarrow 0$ (no self-alignment) the auto-correlation diverges, requiring a different regularization mechanism~\cite{Levine_etal98,Mucha_etal2004}. Figure~\ref{fig:fig2} illustrates another view of the physical mechanism behind the regularization~\footnote{Although this figure seems similar to the one drawn in Ref.~\cite{Koch&Shaqfeh89} for symmetric, rod-like spheroids (which are not self-aligning), the scenarios are different. Whereas the rods are aligned by the flow lines, the ones depicted in Fig.~\ref{fig:fig2}(b) self-align with the force and are only perturbed by the flow lines.}. A concentration fluctuation within a small volume of the suspension creates a flow, which advects objects in and out of the region. Spherical objects respond to the flow isotropically, leading to mutual cancellation of the influx and outflux (Fig.~\ref{fig:fig2}(a)). The dipolar, non-divergenceless flow of irregular objects, as described by Eq.~\eqref{eq:Uv}, perturbs this balance, compensating for the deficiency/surplus of objects in the region (Fig.~\ref{fig:fig2}(b)).

\begin{figure*}
\centerline{\resizebox{0.45\textwidth}{!}{\includegraphics{fig3a.eps}}
\hspace{0.5cm}
\resizebox{0.45\textwidth}{!}{
\includegraphics{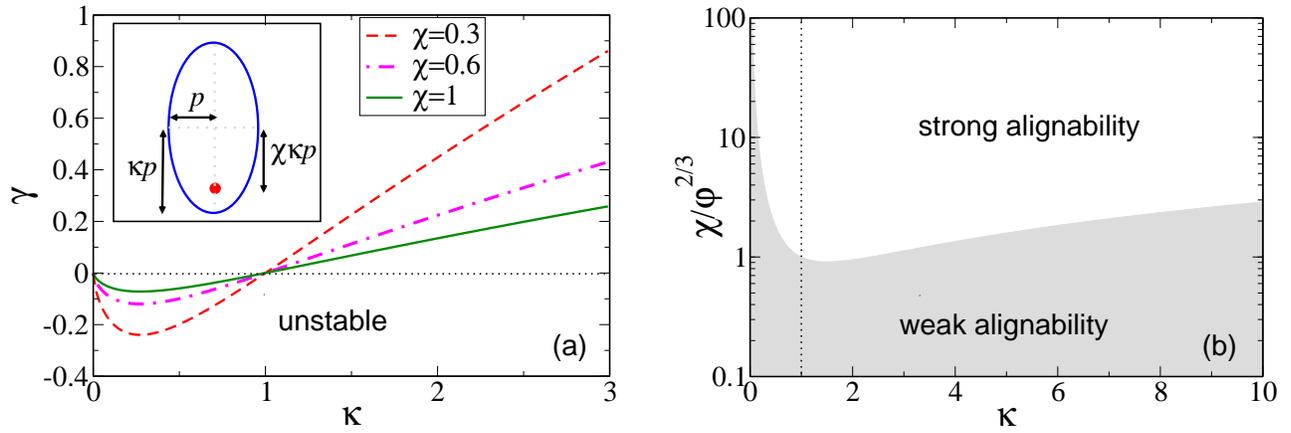}}}
\caption[]{
(a) Effective response parameter $\gamma$ describing suppressed fluctuations for $\gamma>0$ as predicted for sedimentation of self-aligning spheroids (inset). The spheroid's response is calculated as a function of its geometry, characterized by the
aspect ratio $\kappa$, and the displacement of the forcing point (red dot) from the centroid, given by $\chi\kappa p$. Here $p=\kappa^{-1/3}a$ is the spheroid's minor axis. Prolate self-aligning spheroids ($\kappa>1$) cause suppression. For spheres ($\kappa\rightarrow 1)$ our suppression mechanism disappears ($\gamma\rightarrow 0$). For $\kappa<1$ the response parameter $\gamma$ becomes negative so that the uniform suspension becomes unstable.
(b) Validity region of the present theory for self-aligning spheroids. The offset $\chi$ should be larger than $\varphi^{2/3}h(\kappa)$
(where the function $h(\kappa)$ is given in the Supplemental Material~\cite{SM}) to ensure the assumed strong alignability.}  
\label{fig:fig3}
\end{figure*}

As a simple example we treat the specific shape of self-aligning spheroids, i.e., spheroids whose center of mass is displaced from their centroid; inset of Fig.~\ref{fig:fig3}(a). The corresponding response parameter $\gamma$ is shown in Fig.~\ref{fig:fig3}(a) as a function of the spheroid's aspect ratio $\kappa$ and the off-center position of the forcing point $\chi$.  As the offset $\chi$ is reduced, the object is more easily tilted by the flow, thus strengthening the suppression (Fig.~\ref{fig:fig3}(a)). At the same time, however, the object becomes less alignable. Since our calculation is linear in the tilt~\cite{SM}, i.e., it assumes strong alignability, it becomes invalid before the case of a symmetric spheroid ($\chi \rightarrow 0$) is reached. The unshaded area in Fig.~\ref{fig:fig3}(b) indicates this rough validity regime, which involves also the volume fraction $\varphi$. The boundary of this regime gives, roughly, the parameters corresponding to maximum suppression (maximum $\gamma$).

In all of the above we have implicitly assumed that the effective response parameter $\gamma$ is positive, leading to a positive $\xi^2$. As $\gamma\to 0^+$, the characteristic length $\xi$ becomes indefinitely large. In fact, $\gamma$ may be of either sign, as we now show for self-aligning spheroids~\cite{SM}. The response parameter resulting from this calculation, shown in Fig.~\ref{fig:fig3}(a), reveals a region of negative $\gamma$ as a function of the spheroid's aspect ratio $\kappa$.  As is clear from the mechanism described above (Fig.~\ref{fig:fig2}), a negative $\gamma$ implies de-regularization, i.e., instability in the sedimentation of such objects, with unstable structures of size $\sim\sqrt{-\xi^2}$.  The instability clearly calls for additional theoretical and experimental studies.

As described by Eq.~\eqref{eq:Sq} and Fig.~\ref{fig:fig1}(a) (in the case of positive $\gamma$), for any wavevector $\vec{q}\nparallel\hat{\vec{z}}$ the structure factor decays to zero for small wavevectors as $S(\vec{q})\sim q^2$. This implies hyperuniformity of the fluctuating suspension~\cite{Torquato&Stillinger2003,Torquato2016} in any direction but $\hat{\vec{z}}$. Calculating the fluctuation $\delta N$ in the number of objects within a spherical subvolume of radius $R$, we find $\delta N^2 \sim R^3(R/\xi)^{-1}$, i.e., a variance that grows as the surface area rather than the volume~\cite{Torquato2016}. The hyperuniformity is also manifest in the long-range concentration correlations in the transverse direction, as given in Eq.~\eqref{eq:cc}. In the $\hat{\vec{z}}$ direction $S(\vec{q})$ is constant for small $q$, implying normal Poissonian fluctuations. The angular dependence of the suppression has been qualitatively explained above. For $q\neq q_z$ and $q<\xi^{-1}$ the vorticity-tilt effect (with rate $D\xi^{-2}$) dominates diffusion (with the slower rate $Dq^2$), while for $q=q_z$ this effect is absent. Several systems exhibiting hyperuniformity have been recently studied~\cite{Torquato&Stillinger2003,CorteNatPhys2008,WeijsPRL2015,HexnerPRL2015}. Our system is different in several essential aspects: (1) it is dynamic, corresponding to continually changing configurations rather than a static absorbing state, (2) it does not require tuning of a control parameter to a critical value, (3) rather than eliminating collisions, it suppresses both positive and negative concentration fluctuations \footnote{The one-component plasma~\cite{Torquato2016,Ramaswamy2001} is another example of hyperuniformity brought about by long-ranged pair-interactions (electrostatic interactions in this case), falling off as $1/r$.}.

It has been assumed for simplicity that all the objects are identical, but the qualitative conclusions apply in more general scenarios.  The key requirement is that the system contains self-aligning objects, possessing the dipolar anisotropic response treated above. Not all the objects in the suspension need to be self-aligning, and the self-aligning ones do not need to be identical. In any of these scenarios they will tilt and glide in response to the nonuniform flow, thus producing the suppression or instability mechanisms discussed here. 

We now compare our screening mechanisms with the ones suggested for the sedimentation of spheres. Those theories also yielded suppression of fluctuations, but out of different physics. (Indeed, hyperuniformity was found in numerical simulations~\cite{Nguyen&Ladd2005} and experiments~\cite{Lei_etal2001} of sedimenting spheres.) In the theory of Ref.~\cite{Levine_etal98} the effects result self-consistently from the nonlinear coupling between concentration and velocity fluctuations, giving $\xi_{\rm LRFB} \sim a \varphi^{-1/3} \pe^{-2/3}$. A related study~\cite{Ramaswamy2001,Kumar} suggests that the spherical objects might self-organize into structures that glide similarly to our self-aligning objects.
The mechanism of Ref.~\cite{Mucha_etal2004} relies on a steady concentration gradient (stratification), yielding  $\xi_{\rm strat} \sim a \varphi^{-1/4} \pe^{-1/4}$. Which of these is the actual screening mechanism for spheres remains an open question. If the LRFB mechanism is the one that operates for spheres, then, for asymmetric objects, our linear mechanism with $\xi=a\gamma^{-1/2}\varphi^{-1/2} \pe^{-1/2}$ should replace the nonlinear one; the linear glide solution is perturbatively stable against the nonlinear term at steady state~\cite{SM}. If stratification is the active mechanism for spheres, then screening will be caused by a combination of both stratification and asymmetry. Another distinctive feature of the asymmetry mechanism is that it remains in place as the system approaches detailed balance ($D=N$), whereas the LRFB screening disappears~\cite{Levine_etal98}. Thus, we expect the present mechanism to hold for arbitrarily small sedimentation P\'eclet number (while keeping the thermal Pe large).

{\it Conclusion.} The findings presented above for asymmetric dispersions can be checked experimentally, e.g., using light scattering or video microscopy. Our results highlight the different physics underlying the sedimentation of irregular objects as compared to spheroidal ones. This includes a distinctive, direct, screening mechanism, a different length scale $\xi$ beyond which hyperuniformity sets in, and unstable dynamics for certain object shapes. These results may offer new means of controlling the stability of driven suspensions such as fluidized beds.

\begin{acknowledgments}
We thank P. Chaikin, D. Hexner, E. J. Hinch, S. Ramaswamy, and H. Stone for helpful discussions. This research has been supported by the US--Israel Binational Science Foundation
(Grant no.\ 2012090).
\end{acknowledgments}

\bibliography{Sedimentation}

\onecolumngrid

\section{Supplemental Material}

This Supplemental Material contains three parts. In the first one we provide more details on the nature of self-aligning objects and the emergence of an effective, averaged response tensor $\vec{\Phi}$. In the second part we elaborate on the derivation of the different correlation functions presented in the main text. The third part addresses the range of validity of the linear model for the sedimentation of self-aligning objects. 

\subsection{The response tensor $\vec{\Phi}$}

The anisotropic response of self-aligning objects to a weak external flow is captured by the effective hydrodynamic tensor $\vec{\Phi}$ in Eq.~(3). This tensor can be derived from the motion of an isolated object in Stokes flow. Specifically, one should consider the motion of a single rigid object subjected to external force $\vec{F}$, and embedded in external flow gradient $\vec{E}$. Below, we summarize the resulting motion of self-aligning objects according to previous works. In addition, we specialize to a simple example of self-aligning spheroids, for which we derive $\vec{\Phi}$ explicitly.  

\subsubsection{Instantaneous response under external force and flow}
First, we write down the general equations governing the aforementioned hydrodynamic problem. The object is subjected to an external force $\vec{F}=-\hat{\vec{z}}F$, and placed in an external flow $\vec{u}(\vec{r})=\vec{r}^T\cdot\vec{E}$, where $\vec{r}$ is measured from the forcing point, i.e., the point about which the external torque vanishes. We indicate the object's orientation by a set of parameters $\vec{Q}$, e.g., $\vec{Q}$ can refer to three Euler angles or a unit quaternion ~\cite{Goldfriend_etal2015I}. In the inertia-less regime, the instantaneous linear and angular velocities of the object, $\vec{U}$ and $\vec{\Omega}$, are given as a linear response to the external fields~\cite{Kim&Karrila,Goldfriend_etal2015II},
\begin{eqnarray}
\label{eq:SuppU}
U_i &=& -\frac{1}{6\pi\eta a}A_{i3}(\vec{Q})F + a \Pi_{ijk}(\vec{Q})E_{kj}, \\
\label{eq:SuppOm}
\Omega_i &=&-\frac{1}{8\pi\eta a^2}T_{i3}(\vec{Q})F + \left(\Psi_{ijk}(\vec{Q})-\frac{1}{2}\epsilon_{ijk} \right) E_{kj},
\end{eqnarray} 
where $\boldsymbol{\epsilon}$ is the Levi-Civita tensor and the object-dependent tensors $\vec{A}$, $\vec{T}$, $\vec{\Pi}$ and $\vec{\Psi}$ can be calculated by several numerical routines, given the object's shape and mass distribution~\cite{Cortez_etal2005,Torre&Carrasco2002,Filippov2000}.  The velocity $\vec{U}$ in Eq.~\eqref{eq:SuppU} refers to the velocity of the forcing point.  Eq.~\eqref{eq:SuppOm} is a nonlinear equation for the object's orientation (the relation between $\partial_t\vec{Q}$ and $\vec{\Omega}$ is given in Appendix C of Ref.~\cite{Goldfriend_etal2015II} for the case of a unit quaternion representation). The translational motion, described by Eq.~\eqref{eq:SuppU}, varies in time as the object rotates.

\subsubsection{Self-aligning dynamics with $\vec{E}=0$}
 Refs.~\cite{Krapf_etal2009,Gonzalez_etal2004} examined the motion of an isolated self-aligning object in a quiescent fluid. By the definition of these objects, their $\vec{T}$ matrix has only one real eigenvalue. The corresponding eigenvector defines an eigendirection affixed to the object, which will be indicated hereafter by $\hat{\vec{n}}$. The orientational dynamics resulting from Eq.~\eqref{eq:SuppOm} with $\vec{E}=0$ reaches an ultimate state, where $\hat{\vec{n}}$ is aligned with the forcing direction. In addition, the object rotates about $\hat{\vec{n}}$ with a constant rate $\Omega^0=\lambda_0 F$, where $\lambda_0$ is the real eigenvalue of $\vec{T}$. This chiral behavior does not depend on the initial orientation of the object. The ultimate rotation leads to a helical translational motion (dictated by Eq.~\eqref{eq:SuppU}), where on average, the object translates in the direction of the force. The aligned state is also characterized by an aligning rate, $\tau^{-1}_{\rm align}$, which gives the rate to restore alignment from a slightly tilted state~\cite{Krapf_etal2009,Goldfriend_etal2015II}.

\subsubsection{Tilted dynamics with weak $\vec{E}$}
The motion of self-aligning objects in a weak external flow gradient was studied in the context of pair-hydrodynamic interaction~\cite{Goldfriend_etal2015II}. The presence of external flow affects the translational motion directly through the tensor $\vec{\Pi}$, and indirectly due to the orientational motion modified by the tensor $\vec{\Psi}-1/2\boldsymbol{\epsilon}$. When $(\Psi+1/2) E\ll\tau^{-1}_{\rm align}$, the aligning behavior and helical trajectory, corresponding to the case of $\vec{E}=0$, are perturbed. In particular, the object preforms a tilted helical motion, where the averaged eigendirection $\bar{\vec{n}}$ and the averaged linear velocity $\bar{\vec{U}}$, both have components perpendicular to the direction of the force. This perturbed motion can be analyzed by solving Eqs.~\eqref{eq:SuppU} and~\eqref{eq:SuppOm}, where $\vec{\Pi}$ and $\vec{\Psi}$ are constants obtained by averaging over the rotations in the (unperturbed) aligned state. Essentially, the tilted helical dynamics can be characterized as a linear response, i.e., $\bar{n}_i=(\eta a^2/F)\Sigma_{iks}E_{sk}$ and $\bar{U}_i=a\Phi_{iks}E_{sk}$, where $\vec{\Sigma}$ and $\vec{\Phi}$ depend solely on the tensors $\vec{A}$, $\vec{T}$, $\vec{\Pi}$ and $\vec{\Psi}$. We observed such an averaged linear response in the case of self-aligning sparse objects~\cite{Goldfriend_etal2015II}.

\subsubsection{Self-aligning spheroids}

Finally, we specialize to the case of self-aligning spheroids, which can be treated analytically. A self-aligning spheroid is a specific example of a self-aligning object that exhibits a simple motion--- the ultimate translational velocity in the case of $\vec{E}=0$ is constant in time, and does not follow a helical path, i.e., $\lambda_0=0$. 

We consider a spheroid whose principal axes are $(p,p,\kappa p)$, as depicted in the inset of Fig.~3(a) in the main text. 
The typical linear size of the spheroid is related to its volume, $a=\kappa^{1/3}p$.
The spheroid is subjected to an external force at a point which is displaced along the symmetry axis by a distance $\chi\kappa p$ from the center, and embedded in an external flow gradient $\vec{E}$, which is measured about the forcing point. The axisymmetric shape allows us to represent the spheroid's orientation solely by its eigendirection, $\hat{\vec{n}}$, which is directed from the center to the forcing point (indicated by a red dot in Fig.~3). The hydrodynamic tensors in Eqs.~\eqref{eq:SuppU} and~\eqref{eq:SuppOm} refer to the case where the object's {\em origin}--- the point about which the external torque and external flow gradient are measured--- is the forcing point.  For self-aligning spheroids, these tensors are given as a linear transformation from the tensors that correspond to choosing the origin to be the spheroid's center~\cite{Krapf_etal2009,Goldfriend_etal2015II}, which yields:
\begin{eqnarray}
\label{eq:SuppUSph}
U_i &=& -\frac{F}{6\pi\eta p}A^c_{iz}(\hat{\vec{n}})+\chi\kappa p\left(\vec{\Omega}\times\hat{\vec{n}}\right)_i-\chi\kappa p \hat{n}_k E_{ki}, \\
\label{eq:SuppOmSph}
\Omega_i &=&-\frac{\chi\kappa F}{8\pi\eta p^2}S^c_{ij}(\hat{\vec{n}})\epsilon_{jk3}\hat{n}_{k} + \left(\Psi_{ijk}(\hat{\vec{n}})-\frac{1}{2}\epsilon_{ijk} \right) E_{kj}.
\end{eqnarray} 
The tensor $\vec{A}^c$ gives the linear velocity of the spheroid's center as a response to force, and the tensor $\vec{S}^c$ gives its angular velocity as a response to torque about the centeroid. These tensors and the tensor $\vec{\Psi}$ can be found in Ref.~\cite{BrennerIV} and Ref.~\cite{Bretherton1962} respectively. In particular, they are of the form:
$$A^c_{ij}(\hat{\vec{n}})=A_{\perp}(\kappa)\delta_{ij}+\left(A_{\parallel}(\kappa)-A_{\perp}(\kappa)\right)\hat{n}_{i}\hat{n}_{j},$$ 
$$S^c_{ij}(\hat{\vec{n}})=S_{\perp}(\kappa)\delta_{ij}+\left(S_{\parallel}(\kappa)-S_{\perp}(\kappa)\right)\hat{n}_{i}\hat{n}_{j},$$
$$\Psi_{ijk}=-\frac{\psi(\kappa)}{2}\left(\epsilon_{ijm} \hat{n}_m \hat{n}_k  + \epsilon_{ikm} \hat{n}_m \hat{n}_j\right).$$
(Note that Eqs.~\eqref{eq:SuppUSph} and~\eqref{eq:SuppOmSph} follow the description according to  Eqs.~\eqref{eq:SuppU} and~\eqref{eq:SuppOm}. For example, we have $T_{i3}=\chi\kappa S^c_{ij}(\hat{\vec{n}})\epsilon_{jk3}\hat{n}_{k}$.) 

The spheroid's orientational dynamics is given by
\begin{equation}
\Omega_i =-\frac{\alpha F}{\eta p^2}\epsilon_{il3}\hat{n}_{l} + \left(\Psi_{ijk}(\hat{\vec{n}})-\frac{1}{2}\epsilon_{ijk} \right) E_{kj},
\label{eq:SuppOmSph2}
\end{equation} 
where $\alpha=\chi\kappa S_{\perp}(\kappa)/(8\pi)$ is the alignability parameter. 
Eq.~\eqref{eq:SuppOmSph2} leads to the nonlinear equation for the dynamics of eigendirection $\hat{\vec{n}}$,
\begin{equation}
\partial_t\hat{n}_{i}=\epsilon_{iks}\Omega_k \hat{n}_s.
\end{equation}
If $\vec{E}=0$, then the eigendirection and the translational velocity will be ultimately directed along the $-\hat{\vec{z}}$ axis, 
$$
\hat{\vec{n}}^0=-\hat{\vec{z}};\quad \Omega^0=0; \quad U_i^0 = -\frac{F}{6\pi\eta p}A_{\parallel}(\kappa)\delta_{i3}.
$$
This is the self-aligning motion of the isolated object. 
In the case of weak external flow, i.e.,
\begin{equation}
\eta p^2 \psi E \ll \alpha F,
\label{eq:weak}
\end{equation}
the dynamics reaches a state where $\hat{\vec{n}}$ is slightly tilted from the forcing direction, $\hat{\vec{n}}=-\vec{\hat{z}}+\vec{n}^{\perp}$. The asymptotic solution can be found by setting $\vec{\Omega}=0$ in Eq.~\eqref{eq:SuppOmSph2} and approximating $\Psi_{iks}(\hat{\vec{n}})=\Psi_{iks}(-\hat{\vec{z}})$, which gives
\begin{equation}
\epsilon_{il3} n^{\perp}_{l} =
\frac{\eta p^2}{\alpha F} \left(\Psi_{ijk}(-\hat{\vec{z}})-\frac{1}{2}\epsilon_{ijk} \right) E_{kj},
\label{eq:rot1}
\end{equation}
or
\begin{equation}
n^{\perp}_{i} =
\frac{\eta p^2}{\alpha F} \epsilon_{i3m}\left(\Psi_{mjk}(-\hat{\vec{z}})-\frac{1}{2}\epsilon_{mjk} \right) E_{kj}.
\label{eq:rot2}
\end{equation}
 
Next, let us consider the perturbed translational velocity in Eq.~\eqref{eq:SuppUSph}. Up to linear order in $E$ we have 
\begin{equation}
 U_i-U^0_i=\frac{F}{6\pi \eta p} \left(A_{\parallel}(\kappa)-A_{\perp}(\kappa)\right)n^{\perp}_i+p\chi\kappa \hat{z}_k E_{ki}.
\label{eq:lin1}
\end{equation}
Finally, substituting in Eq.~\eqref{eq:lin1} the final orientation given in Eq.~\eqref{eq:rot2}, we find
\begin{equation}
 U_i-U^0_i=\frac{p\beta}{\alpha} \epsilon_{i3m}\left(\Psi_{mjk}(-\hat{\vec{z}})-\frac{1}{2}\epsilon_{mjk} \right) E_{kj}+\chi\kappa p\hat{z}_k E_{ki} + O(E^2),
\label{eq:tran2}
\end{equation}
where $\beta=\left(A_{\parallel}(\kappa)-A_{\perp}(\kappa)\right)/(6\pi)$ is the gliding parameter of the object.
To summarize, recalling the relation $a=\kappa^{1/3}p$, we have $U_i-U^0_i= a \Phi_{ijk}E_{kj} + O(E^2)$ with
\begin{equation}
\Phi_{ijk}=\chi\kappa^{2/3} \delta_{ij}\delta_{3k}+\frac{\beta \kappa^{-1/3}}{\alpha} \epsilon_{i3m}\left(\Psi_{mjk}(-\hat{\vec{z}})+\frac{1}{2}\epsilon_{mjk} \right) .
\end{equation}

Here we provide the following explicit expressions for the different hydrodynamic parameters, as well as their asymptotic values for an extremely oblate spheroid ($\kappa\ll 1$), a sphere ($\kappa=1$), and an extremely prolate spheroid ($\kappa\gg 1$): 
\begin{equation}
\alpha(\kappa,\chi) \equiv \chi\kappa S_{\perp}(\kappa)/(8\pi)= 
3\chi \kappa
	\frac{\kappa\sqrt{1-\kappa^2}+(1-2\kappa^2)C^{-1}(\kappa)}{16\pi (1+\kappa^2)(1-\kappa^2)^{3/2}} 
= \left\{ 
\begin{array}{l l}
	\frac{3 \chi \kappa}{32} & ,\kappa \ll 1 \\
	\frac{\chi}{8\pi}   & ,\kappa=1\\
	\frac{3\chi\ln(\kappa)}{8\pi\kappa^2} & , \kappa \gg 1
\end{array}
\right. ,
\label{eq:lam}
\end{equation}
\begin{equation}
\beta(\kappa) \equiv \frac{A_{\parallel}(\kappa)-A_{\perp}(\kappa)}{6\pi} = 
\frac{3\kappa\sqrt{1-\kappa^2}-(1+2\kappa^2)C^{-1}(\kappa)}{16\pi (1 - \kappa^2)^{3/2}}
=\left\{ 
\begin{array}{l l}
	-\frac{1}{32} & ,\kappa \ll 1 \\
	0   & ,\kappa=1\\
	\frac{\ln(\kappa)}{8\pi\kappa} & , \kappa \gg 1
\end{array}
\right. ,
\label{eq:bet}
\end{equation}
\begin{equation}
\gamma(\kappa,\chi)\equiv \Phi_{zzz}-\Phi_{z\perp\perp}-\Phi_{\perp\perp z}=
\frac{\beta(\kappa)}{\alpha(\kappa,\chi)}\frac{\kappa^{5/3}}{1+\kappa^2}=
\left\{ 
\begin{array}{l l}
	-\frac{\kappa^{2/3}}{3\chi} & ,\kappa \ll 1 \\
	0 & , \kappa=1\\
	\frac{\kappa^{2/3}}{3\chi} & , \kappa \gg 1
\end{array}
\right. ,
\end{equation}
where 
$$C^{-1}(x)=\left\{ 
\begin{array}{l l}
	\cos^{-1}(x) & ,x \leq 1 \\
	i\cosh^{-1}(x) & , x>1
\end{array}
\right. .
$$

For extremely oblate or prolate spheroids, in the limit of $\chi\to 0$ one finds that $|\gamma|\rightarrow\infty$; however, in this limit $\alpha\to 0$ and the assumption of weak external flow gradient in Eq.~\eqref{eq:weak} breaks.
This assumption can be related to the volume fraction as follows. The external flow around the object is given by the Oseen flow generated by other forced objects in the system, $u\sim F/(8 \pi \eta l)$, where $l$ is the typical distance between objects. Thus, the flow gradient is given by $E\sim F/(8 \pi \eta l^2)$, and the inequality in Eq.~\eqref{eq:weak} reads 
$\kappa^{-2/3}\varphi^{2/3}\ll 8\pi\alpha=\chi\kappa S_{\perp}(\kappa)$, or
\begin{equation}
\chi\gg h(\kappa)\varphi^{2/3} \equiv\frac{\kappa^{-5/3}}{S_{\perp}(\kappa)}\varphi^{2/3}, 
\label{eq:weakE}
\end{equation}
where $\varphi=(a/l)^{1/3}$.
Eq.~\eqref{eq:weakE} defines the regime of strong alignability which appears in Fig.~3(b) in the main text.

\subsection{Real space correlations}

Below we summarize the detailed calculations yielding the real space correlations that are given in the text.

\subsubsection{Total and direct density-density correlation functions}
The density-density correlations are the inverse Fourier transform of the suspension's static structure factor:
$$
\langle c(0)c(\vec{r}) \rangle=c_0
 \int{\frac{d^3\vec{q}}{(2\pi)^3} S(\vec{q}) e^{-i \vec{q}\cdot\vec{r}}}=
\frac{Nc_0}{D}\int \frac{d^3\vec{q}}{(2\pi)^3} \frac{q^2 e^{-i \vec{q}\cdot\vec{r}}}{q^2+\xi^{-2}(q_{\perp}/q)^2}.
$$
We rescale $r\rightarrow \xi R$ and define the rescaled total correlation function 
\begin{equation}
h(\vec{R})\equiv
\frac{D\xi^3}{N c_0} \langle c(0)c(\vec{R}) \rangle=
\int \frac{d^3\vec{q}}{(2\pi)^3} \frac{q^2 e^{-i \vec{q}\cdot\vec{R}}}{q^2+(q_{\perp}/q)^2}.
\end{equation}
We start with correlations along the direction of gravity
\begin{equation}
h(R\hat{\vec{z}})=
\int \frac{dq_{\perp}dq_z }{(2\pi)^2}   
\frac{(q_z^2+q^2_{\perp})^2  q_{\perp} e^{-i q_z R} }{(q_z^2+q^2_{\perp})^2 +q^2_{\perp}}.
\label{eq:hz}
\end{equation}
Noticing that
\begin{equation}
\int{  
\frac{q_{\perp} e^{-i q_z R}}{(q_z^2+q^2_{\perp})^2 +q^2_{\perp}} 
dq_z }= -\pi \text{Im}\left(\frac{e^{-\sqrt{q^2_{\perp}+iq_{\perp}}R}}{\sqrt{q^2_{\perp}+iq_{\perp}}}\right),
\end{equation}
we find
\begin{equation}
h(R\hat{\vec{z}})=-  
\text{Im} \left(
\int \frac{dq_{\perp}}{4\pi} \left[ (q^2_{\perp}+iq_{\perp})^{3/2} 
-2q^2_{\perp}\sqrt{q^2_{\perp}+iq_{\perp}}
+q^4_{\perp} (q^2_{\perp}+iq_{\perp})^{-1/2} \right] 
 e^{-\sqrt{q^2_{\perp}+iq_{\perp}}R}
\right).
\end{equation}
In the limit $R\to \infty$, the first term is dominant over the others, which gives
\begin{equation}
\lim_{R\to \infty} h(R\hat{\vec{z}})=
-\text{Im}\left(  
\int{(iq_{\perp})^{3/2}e^{-\sqrt{iq_{\perp}}R}\frac{dq_{\perp}}{4\pi}} \right)=
\frac{12}{\pi R^5}.
\end{equation}

Next, consider correlations in the direction perpendicular to gravity
\begin{equation}
h(\vec{R}_{\perp}) =
\int  \frac{dq_z dq_{\perp}d\phi }{(2\pi)^3}
\frac{ (q_z^2+q^2_{\perp})^2  q_{\perp}e^{-i q_{\perp} R \cos(\phi)} }{(q_z^2+q^2_{\perp})^2 +q^2_{\perp}} 
  =
\int \frac{dq_{\perp}dq_z}{(2\pi)^2}  
\frac{(q_z^2+q^2_{\perp})^2 q_{\perp} }{(q_z^2+q^2_{\perp})^2 +q^2_{\perp}} J_0(q_{\perp} R),
\label{eq:hperp}
\end{equation}
where $J_0$ is the Bessel function of the first kind.
In order to preform the following calculation we use the relations:
\begin{equation}
\frac{q_{\perp}}{(q_z^2+q^2_{\perp})^2 +q^2_{\perp}} =
\text{Re}\left(\frac{1}{q_{\perp}-i f_{-}(q_z)}-
\frac{1}{q_{\perp}+i f_{+}(q_z)}\right)
\frac{1}{f_{+}(q_z)+f_{-}(q_z)},
\end{equation}
with $f_{\pm}(q_z)=\frac{\sqrt{1+4q^2_z}\pm 1}{2}$, and  
\begin{equation}
\text{Re}\left( \int{J_0(q_{\perp}R) \frac{dq_{\perp}}{q_{\perp}-i b}}\right)=K_0(bR),
\end{equation}
where $K_0$ is the modified Bessel function of the second kind. Integration over $q_{\perp}$ in Eq.~\eqref{eq:hperp} reads
$$
h(\vec{R}_{\perp})=M_1+M_2+M_3, 
$$
where
\begin{eqnarray}
M_1&=&
\int{\frac{dq_z}{(2\pi)^2} \frac{q^4_z}{f_{+}(q_z)+f_{-}(q_z)}\left [ K_0(f_{-}(q_z)R) -K_0(f_{+}(q_z)R) \right]}, \\
M_2&=&
-\int{\frac{dq_z}{(2\pi)^2}\frac{2q^2_z}{f_{+}(q_z)+f_{-}(q_z)}\left [f^2_{-}(q_z) K_0(f_{-}(q_z)R) -
f^2_{+}(q_z) K_0(f_{+}(q_z)R) \right]},\\
M_3&=&
\int{\frac{dq_z}{(2\pi)^2} \frac{1}{f_{+}(q_z)+f_{-}(q_z)}\left [ f^4_{-}(q_z) K_0(f_{-}(q_z)R) -
f^4_{+}(q_z) K_0(f_{+}(q_z)R) \right]}. 
\end{eqnarray}
In the limit of $R\to \infty$, the first term dictates the leading order behavior
$$
\lim_{R\to \infty} h(\vec{R}_{\perp})= 
\int{\frac{dq_z}{(2\pi)^2}   q^4_z K_0(q_z^2 R) } =
\frac{\Gamma^2(5/4)}{2 \sqrt{2} \pi^2 R^{5/2}}.
$$

Finally, we calculate the direct correlation function, $d(\vec{R})$, that can be written explicitly, and not in an integral form, as opposed to the total correlation function given above. The two functions are related through the Ornstein-Zernike equation, which implies~\cite{Torquato&Stillinger2003}
$$
\tilde{d}(\vec{q})=
 \frac{S(\vec{q})-1}{c_0S(\vec{q})}.
$$
In the case of $N=D$ we have
$$
\tilde{d}(\vec{q})=
 -\frac{\xi^{-2}q^2_{\perp}}{c_0 q^4},
$$
and returning back to real space we get
\begin{equation}
d(\vec{r})=-\frac{1}{c_0\xi^{2}} 
\int \frac{d^3\vec{q}}{(2\pi)^3} \frac{q^2_{\perp} e^{-i\vec{q}\cdot\vec{r}} }{ q^4}   =
-\frac{1}{8\pi c_0 \xi^{2}  r} \left( 1+\frac{r^2_{z}}{r^2} \right).
\end{equation}

\subsubsection{Velocity Correlations}

Let us now calculate the velocity two-point correlation function
\begin{equation}
\langle U_i(0)U_j(\vec{R})\rangle
= \frac{N F^2\xi}{D} \int\frac{dq^3}{(2\pi^3)} 
\frac{\tilde{G}_{iz}(\vec{q}) \tilde{G}_{jz}(-\vec{q}) q^2 e^{-i\vec{q}\cdot\vec{R}}}
{q^2+(q_{\perp}/q)^2}.
\label{eq:UUqSupp}
\end{equation}
Separating to the cases of correlations along- and perpendicular to the direction of gravity we get the following expressions:

\begin{equation}
\langle U_z(0)U_z(\vec{R})\rangle=
\frac{N F^2\xi}{D}\int
\frac{d^3\vec{q}}{(2\pi)^3} \frac{q^2_{\perp}}{q^4}
 \frac{q^2_{\perp}e^{-i \vec{q}\cdot\vec{R}}}{q^4 +q^2_{\perp}} ,
\end{equation}
\begin{equation}
\langle U_{\perp}(0)U_{\perp}(\vec{R})\rangle=
\frac{\langle U_x(0)U_x(\vec{R})\rangle+\langle U_y(0)U_y(\vec{R})\rangle}{2}=
\frac{N F^2\xi}{2D} \int \frac{d^3\vec{q}}{(2\pi)^3}
\frac{q_z^2}{q^4}
 \frac{q^2_{\perp} e^{-i \vec{q}\cdot\vec{R}}}{q^4 +q^2_{\perp}} .
\end{equation}
Note that
$$
\frac{1}{q^4}
 \frac{q^2_{\perp}}{q^4 +q^2_{\perp}}=
\frac{1}{q^4}-\frac{1}{q^4 +q^2_{\perp}},
$$
thus, the integrals become
\begin{equation}
\langle U_z(0)U_z(\vec{R})\rangle=
\frac{N F^2\xi}{D}  \left(
\int{ \frac{d^3\vec{q}}{(2\pi)^3}
\frac{q^2_{\perp}}{q^4} 
e^{-i \vec{q}\cdot\vec{R}}}-
 \int{ \frac{d^3\vec{q}}{(2\pi)^3}
 \frac{q^2_{\perp}}{q^4 +q^2_{\perp}} 
e^{-i \vec{q}\cdot\vec{R}}
} \right),
\end{equation}
\begin{equation}
\langle U_{\perp}(0)U_{\perp}(\vec{R})\rangle=
\frac{N F^2\xi}{2D} \left(
\int{\frac{d^3\vec{q}}{(2\pi)^3}
\frac{q^2_{z}}{q^4} 
e^{-i \vec{q}\cdot\vec{R}}}-
 \int{\frac{d^3\vec{q}}{(2\pi)^3}
 \frac{q^2_{z}}{q^4 +q^2_{\perp}} 
e^{-i \vec{q}\cdot\vec{R}}
} \right).
\end{equation}
The first integral in the two expressions above can be solved explicitly
\begin{equation}
\int{\frac{d^3\vec{q}}{(2\pi)^3}  
\frac{q^2_{\perp}}{q^4} 
e^{-i \vec{q}\cdot\vec{R}}} = \frac{1}{8\pi R} \left( 1+\frac{R^2_{z}}{R^2} \right),
\end{equation}
\begin{equation}
\frac{1}{2} 
\int{\frac{d^3\vec{q}}{(2\pi)^3} 
\frac{q^2_{z}}{q^4} 
e^{-i \vec{q}\cdot\vec{R}}} = \frac{1}{16\pi R} \frac{R^2_{\perp}}{R^2}.
\end{equation}
Finally, the velocity correlations along the longitudinal and perpendicular directions can be found similarly to the integrals calculated above:
\begin{eqnarray}
\frac{D\langle U_z(0)U_z(R\hat{\vec{z}})\rangle}{N F^2\xi}&=&
\frac{1}{4\pi R}
+\pi \text{Im}\left(\int \frac{dq_{\perp}}{(2\pi)^2}
 \frac{q^2_{\perp} e^{-\sqrt{q^2_{\perp}+iq_{\perp}}R}}{\sqrt{q^2_{\perp}+iq_{\perp}}}\right)
\\
\frac{D\langle U_{\perp}(0)U_{\perp}(R\hat{\vec{z}})\rangle}{N F^2\xi}&=&
-\frac{\pi}{2} \text{Im}\left(\int\frac{dq_{\perp}}{(2\pi)^2}   \sqrt{q^2_{\perp}+iq_{\perp}} e^{-\sqrt{q^2_{\perp}+iq_{\perp}}R} \right)
\\
\frac{D\langle U_z(0)U_z(\vec{R}_{\perp})\rangle}{N F^2\xi}&=&
 \frac{1}{8\pi R}+
\int \frac{dq_z}{(2\pi)^2}
\frac{f^2_{-}(q_z)K_0(f_{-}(q_z)R)-f^2_{+}(q_z)K_0(f_{+}(q_z)R)}{f_{+}(q_z)+f_{-}(q_z)} 
\\
\frac{D\langle U_{\perp}(0)U_{\perp}(\vec{R}_{\perp})\rangle}{N F^2\xi}&=&
\frac{1}{16\pi R}-\frac{1}{2}
\int
\frac{dq_z}{(2\pi)^2} \frac{q^2_z\left ( K_0(f_{-}(q_z)R) -K_0(f_{+}(q_z)R) \right)}
{f_{-}(q_z)+f_{+}(q_z)} 
\end{eqnarray}

\subsection{Validity of Linear theory}

The original fluctuating hydrodynamics model, Eq. (4), contains an advective term which is linear in $c$, and a nonlinear one, $\sim c^2$, which underlies the LRFB model. The ratio between the two terms reads    
\begin{equation}
\frac{\text{anisotropic linear term}}{\text{LRFB nonlinear term}}=
\frac{a \gamma  \tilde{G}(q) F q^{-1} \tilde{c} c_0}
{\tilde{G}(q) F q \tilde{c}\tilde{c}}=
\gamma \varphi (aq)^{-2} \frac{1}{\tilde{c}}.
\label{eq:eqr}
\end{equation}
The concentration fluctuations appearing in Eq.~\eqref{eq:eqr} can be related to the static structure factor
\begin{equation}
\tilde{c} \sim \sqrt{S}=\sqrt{ N/D} \sqrt{\frac{1}{1+(\xi q)^{-2}}} \underset{q\xi\ll 1}{\rightarrow}
\sqrt{ N/D}q\xi .
\label{eq:ccSupp}
\end{equation}
The ratio in Eq.~\eqref{eq:eqr} is then proportional to $(q\xi)^{-3}$, and thus, much larger than 1 as $q\to 0$. Therefore, the linear term in Eq.~(4) dominates
the nonlinear one in both limits of small concentration fluctuations and small $q$. In these limits, the linear theory described by Eq.~(5) is valid.

\end{document}